\documentclass[12pt]{article}
\usepackage{a4,amsmath,amsxtra,amssymb}

\begin{document}

\baselineskip=1.5\baselineskip

\centerline{\textbf{\Large An Unexpected Electrovac Solution }}

\centerline{\textbf{\Large with the Negative Cosmological Constant
}}

\vglue16pt

\centerline{\bf Jerzy Klemens Kowalczy\'nski}

\centerline{Institute of Physics, Polish Academy of Sciences,}

\centerline{Al. Lotnik\'ow 32/46, 02-668 Warsaw, Poland}

\centerline{E-mail address: jkowal@ifpan.edu.pl}

\vglue24pt\noindent
{\bf Abstract.} An exact solution of the current-free
Einstein--Maxwell
equations with the cosmological constant is presented. The
solution is of Petrov type D, includes the negative cosmological
constant, and could be a ``background addition'' to the
present-day models of the universe. It has a surprising property
such that its electromagnetic field and cosmological constant are
interdependent (this constant is proportional to the energy
density of this field), which may suggest a new way of measuring
the constant in question. The solution describes a constant
electromagnetic background with a preferred direction in the
universe, and defines the entire lifetime of the universe as a
simple function of the negative cosmological constant. According
to our solution the absolute value of this constant should be
considerably lower than that recently estimated, when
astrophysical data are taken into account. Our solution is a special
case of that published by Bertotti in 1959. His solution (in terms
of which the cosmological constant and the background
electromagnetic field are independent) and its two other special
cases, i.e. the conformally flat Robinson solution (1959) and the
one which is the counterpart of our solution with the positive
cosmological constant, are briefly discussed.

\vglue20pt\noindent

\eject
The present-day descriptions of the universe are an
important domain of general relativity. Apart from several
well-known homogeneous models of the universe, there is a large
number of inhomogeneous ones [1]. Both classes are considered with
or without the cosmological constant, $ \Lambda $, which has
advocates and opponents. Some advocates even consider this
constant to be inevitable in the models [2]. The question of $
\Lambda $ in cosmology is presented in detail by Carroll, Press,
and Turner in their review article [2] provided with a huge list
of topical literature. The effects of the cosmological constant on
the homogeneous models are also described in Ref. [3] with lucid
and concise summaries on pp. 746, 747, 773, and 774. Examples of
the strong effect in the case of inhomogeneous models are
presented on pp. 25 and 27 in Ref. [1]. In general, the models
with and without $ \Lambda $ may considerably differ.

The problem of whether the cosmological constant exists is
therefore one of the central problems of cosmology today. In order
to settle it, astrophysicists implant their observational data
(mean mass-energy density of the universe, deceleration parameter,
Hubble ``constant'') in equations resulting from the so-called
cosmological solutions of the Einstein equations for dust or
fluid. However, there is considerable discrepancy between the
observational data (see, e.g., Refs. [3,4] and p. 587 in Ref.
[5]). Besides, there is no consensus as to which one of those
solutions (i.e. cosmological models) should be taken into account
as the best approximation to reality. The possible values of $
\Lambda $, positive or negative, are, therefore, only roughly
estimated. The recent estimation [6], based on Refs. [2,4], is
$$
 0 \leq |\Lambda| \lesssim 2.2 \times 10^{-56}\
{\rm cm}^{-2}\eqno {\rm(1)}
$$
in the case of $ \Lambda \leq 0 $,
which is the one we are interested in here. (For $ \Lambda \geq 0
$ the upper limit is two times higher [4,6]). There is no lower
limit different from zero, i.e. the models with $ \Lambda = 0 $
are not excluded.

In consequence, today the estimation of $ \Lambda $ is not only
complicated but also uncertain and indirect. It appears, however,
that the Einstein--Maxwell theory points to a different approach,
possibly simpler and more direct, since it consists in measuring a
constant electromagnetic background of the universe. Such a
possibility is illustrated below for $ \Lambda \leq 0$, i.e. for
the case when the presence of $ \Lambda $ decelerates the
expansion of the universe.

The following simple metric form
$$
 ds^{2} = dx^{2} + dy^{2} + 2\left(1 + \Lambda uv\right)^{-2}\,du\,dv
 \eqno {\rm (2)}
$$
and electromagnetic field tensor
$$ F_{xy} = p,\quad F_{uv} =
\left(1 + \Lambda uv\right)^{-2}q,\quad F_{xu} = F_{xv} = F_{yu} =
F_{yv} = 0,\eqno {\rm (3)}
$$
with
$$
 \Lambda = -c^{-4}G\left(p^{2} + q^{2}\right), \eqno{\rm(4)}
$$
where $ p $ and $ q $ are real constants, $ c $ is the speed of
light in vacuum, and $ G $ is the Newtonian gravitational
constant, are an exact solution of the current-free
Einstein--Maxwell equations with the cosmological constant
$$
 R_{\mu\nu} - \textstyle{1\over2}g_{\mu\nu}R = g_{\mu\nu}\Lambda +
 2c^{-4}G\left(F_{\sigma\mu}F_{\nu}\vphantom{F}^{\sigma} +
 \textstyle{1\over4}
 g_{\mu\nu}F_{\sigma\tau}F^{\sigma\tau}\right),
$$
$$
 F_{\left[\mu\nu,\sigma\right]} = 0,\qquad F^{\mu\nu}
 \vphantom{F}_{;\nu} = 0,
 \eqno {\rm (5)}
$$
and where the signature ${} + + + -{} $ and convention $
R_{\mu\nu}
:= R^{\sigma}\vphantom{R}_{\mu\nu\sigma} $ are assumed.
The invariant and pseudoinvariant of the electromagnetic field are
$$ F_{\mu\nu}F^{\mu\nu} = 2\left({\bf B}^{2} - {\bf E}^{2}\right)
= 2\left(p^{2} - q^{2}\right),\quad
F_{\mu\nu}\widetilde{F}^{\mu\nu} = 4{\bf E}{\bf B} = 4pq,
\eqno{\rm(6)} $$ where $ \widetilde{F}^{\mu\nu} $ is the dual of $
F_{\mu\nu} $, and $ {\bf E} $ and $ {\bf B} $ are three-vectors of
the electric and magnetic fields, respectively. Thus
our electromagnetic field is non-null.
Metric (2) is of Petrov type D. Its two Debever--Penrose vectors $
k^{\mu} $ and $ l^{\mu} $, each double of course, can have the
covariant components $ k_{\mu} = \delta_{\mu}^{u} $ and $ l_{\mu}
= \delta_{\mu}^{v} $; they are geodesic, shear-free,
rotation-free, and expansion-free. These are also principal null
vectors of our electromagnetic field, i.e. we have here a doubly
aligned case. From Eq. (4) we see that $ \Lambda \leq 0$
in our solution.

The general form of our solution is $ ds^{2} = dx^{2} + dy^{2} +
2e^{A}\,du\,dv$, where a disposable function $ A =
A\left(u,v\right) $ is restricted by the condition $ e^{-A}A_{,uv}
= -2\Lambda $, but owing to this condition we can retransform $ u
$ and $ v $ so as to get rid of the disposable function and obtain
the metric (2). For a proof see Refs. [7,8].

After making a coordinate transformation
$$
x = x,\quad y = y,\quad u = 2^{1/2}j^{-1}e^{jz}M,\quad v =
-2^{1/2}j^{-1}e^{-jz}M,
$$
$$
 j := \left(-2\Lambda\right)^{1/2},\qquad M := \tan
 \left(\textstyle{1\over2} jct + \textstyle{1\over4}\pi\right),
 \eqno {\rm(7)}
$$
whose Jacobian is
$$
 {\partial\left(x,y,u,v\right)\over\partial\left(x,y,z,t\right)}
 = -4c\left(1 - \sin jct\right)^{-2}\cos jct,\eqno {\rm(8)}
$$
we get our metric (2) in synchronous coordinates
$$
 ds^{2} = dx^{2} + dy^{2} + \cos ^{2}\left(jct\right)\,dz^{2} -
 c^{2}dt^{2}, \eqno {\rm (9)}
$$ i.e. $ t $ is the cosmic time; and the Cartesian-like
components of $ {\bf E} $ and $ {\bf B} $ are $$ E_{x} = E_{y}=
0,\qquad E_{z} = q = \pm|{\bf E}|, $$ $$ B_{x} = B_{y} = 0,\qquad
B_{z} = p = \pm|{\bf B}|.\eqno {\rm (10)} $$ Note that the
limiting transition $ \Lambda \to 0 $ does not make the
determinant (8) singular.

In all known to me exact solutions of Eqs. (5) the cosmological
constant and electromagnetic field are independent.
(Note added: there exists a counterpart of our solution,
see the end of the present article.) Solution
(2)--(4) is therefore a surprise. In virtue of Eqs. (4) and (6)
the existence of fields (10) is {\it equivalent\/} to the
existence of a negative $ \Lambda $. The fields (10) constitute a
{\it constant electromagnetic background\/} in the whole universe
(if and only if $ \Lambda < 0 $). This background is not uniquely
determined by the value of $ \Lambda $ since $ p $ and $ q $ are
independent. For every given $ \Lambda < 0 $ we can have, by Eqs.
(10), arbitrary values of $ {\bf E}^{2} $ and $ {\bf B}^{2} $
within  Eq. (4), including the extreme cases $ \Lambda =
-c^{-4}G{\bf E}^{2} $ for $ {\bf B} = 0 $ and $ \Lambda =
-c^{-4}G{\bf B}^{2} $ for $ {\bf E} = 0 $. It is seen from Eqs.
(10) that $ {\bf E} $ and $ {\bf B} $ are parallel if $ {\bf
E}{\bf B} \neq 0 $ (or antiparallel; none of Eqs. (3)--(6)
determines the signs of $ p $ and {\it q\/}). Thus $ {\bf E} $ or
$ {\bf B} $ (both if $ {\bf E}{\bf B} \neq 0 $) determines a {\it
physically preferred\/} spacelike direction in the universe.

If one assumes that our solution describes a physical (cosmic)
reality, then one admits a simple and almost {\it direct\/} method
of measuring the negative cosmological constant, by searching for
and measuring the electromagnetic background. If none of the
fields (10) is discovered, then one concludes that the absolute
value of $ \Lambda < 0 $ lies below the sensitivity threshold of
the measurement, and the possibility of $ \Lambda \geq 0 $ is
admitted. There are, however, Maxwellian plasmas in the
interstellar and intergalactic spaces [5]. We have to assume
therefore that $ |{\bf E}| $ is extremely small or even zero since
otherwise such plasmas could not exist. Then Eqs. (4) and (10)
give $$
 \Lambda \cong -c^{-4}G{\bf B}^{2}.\eqno {\rm (11)}
$$

The proposed method of measuring the negative $ \Lambda $ is
simple in principle though not necessarily in practice. In our
cosmic neighbourhood we have various complicated structures of
relatively strong magnetic fields [9--11], which can conceal the
presumable background magnetic field. The background should
therefore be sought in the large-scale extragalactic space.
Unfortunately, in this wider scale we have an analogous situation,
though the structures are, in general, considerably larger [12,13]
and the fields considerably weaker [5,14]. Lemoine {\it et al.\/}
[14] estimate the extragalactic magnetic field strength at $ \sim
1 $ pG -- $ \sim 1 $ nG, however, this concerns the
root-mean-squared strength (denoted by $ B_{\rm{rms}} $ in Ref.
[14]). This quantity is used from necessity because in general we
do not know the directions and senses of almost constant magnetic
fields occurring in large extragalactic regions of space [12]. $
B_{\rm{rms}} $ includes, by definition, uncertainties and
therefore cannot be directly related to our background field $
{\bf B} $. For instance, if we perform observations through the
large regions just mentioned, and if the senses of magnetic fields
in these regions differ, then the measured $ B_{\rm{rms}} $ may be
considerably lower than the field strength in some of the regions.
Nevertheless, it seems obvious that the strength $ |{\bf B}| $ of
our background cannot be much higher than the values estimated
above. Assuming tentatively $ |{\bf B}| \lesssim 10 $ nG, from
relation (11) we get $$
 |\Lambda| \lesssim 8.3 \times 10^{-66}\ {\rm cm}^{-2},
 \eqno {\rm (12)}
$$
i.e. values considerably lower than the upper limit in relation
(1), and so small that one might even doubt whether $ \Lambda < 0
$ exists at all (cf. remark in reference 18 on p. 71 in Ref. [6]).
On the other hand, we may not {\it a priori\/} exclude the
existence of the background, but the relevant settlement would
need laborious observations which should take into account the
possibility of the background field occurring.

Our solution is of course too simple to represent a model of the
universe. It can just be a ``background addition'' to the
present-day models based on dust or fluid type solutions. For
instance, if one assumes an isotropic model, then our solution can
introduce an anisotropic perturbation (preferred direction); the
lower the $ |\Lambda| $ the smaller the deviation from isotropy
(relation (11)).

Let us still note another aspect of our solution. If $ \Lambda < 0
$, then metric (9) degenerates for $ t = T_{n} := \left(n +
{1\over2}\right)\pi/jc $ and $ n = 0 $, $ \pm1 $, $ \pm2 $, \dots
(one space dimension vanishes). Thus each $ T_{n} $ may be
interpreted as an instant of the death of a universe and of the
birth of the next one. The solution suggests therefore the
existence of an infinite sequence of universes; and then a period
$$
 T := T_{n + 1} - T_{n} = \pi/jc \eqno {\rm(13)}
$$
would be the entire lifetime (from birth to death) of each of
them. Relations (12) and (13) give $ T \gtrsim 0.8 $ Pyr, i.e. a
period almost $ 10^{5} $ times longer than the recent estimation
[6] of the age of our universe!

{\bf Note added.} Our solution is a special case of that found by
Bertotti [15]. His solution also describes a spacetime with constant
electric ($ {\bf E} $) and magnetic ($ {\bf B} $)
fields that also are independent and parallel or antiparallel
(if $ {\bf E}{\bf B} \neq 0 $). When the coordinate system of
Eq. (9) is used the Bertotti solution takes the form
$$
 ds^{2} = dx^{2} + \cos ^{2}\left(ax\right)\,dy^{2} + 
\cos ^{2}\left(bct\right)\,dz^{2} -
 c^{2}dt^{2}, \eqno {\rm (14)}
$$ where the real constants $a$ and $b$ are determined by equations
$$
 a^{2} = c^{-4}G\left({\bf B}^{2} + {\bf E}^{2}\right) + \Lambda
$$
$$
 b^{2} = c^{-4}G\left({\bf B}^{2} + {\bf E}^{2}\right) - \Lambda,
 \eqno{\rm(15)}
$$
and the axis $z$ is then parallel to the three-vectors
$ {\bf E} $ and $ {\bf B} $. In this case, unlike our solution
($a = 0$), the energy density of the background electromagnetic
field is independent of $ \Lambda $. The metric form (14) is of
Petrov type D iff $a^{2} \neq b^{2}$ (i.e. iff $ \Lambda \neq 0$ for
the solution (14) and (15)). Its special case
$a^{2} = b^{2}$ (i.e. iff $ \Lambda = 0$) is conformally flat and is
called the Robinson solution [16]. Another special case, $a \neq 0$
and $b = 0$, is also an exact solution of the current-free
Einstein--Maxwell equations. This solution is of Petrov type D and
has properties analogous to those of our solution but with
$ \Lambda > 0 $ and periodicity in space.

\vskip20pt

I wish to thank S. Bajtlik, M. Bzowski, and B. Mielnik for helpful
discussions.

\vskip20pt
\centerline{{\bf References}}

\vglue12pt

[1] A. Krasi\'nski, {\it Inhomogeneous Cosmological Models\/}
(Cambridge University Press, Cambridge, 1997).

[2] S.M. Carroll, W.H. Press, and E.L. Turner, Annu. Rev. Astron.
Astrophys. {\bf30}, 499 (1992).

[3] C.W. Misner, K.S. Thorne, and J.A. Wheeler, {\it
Gravitation\/} (Freeman and Co., San Francisco, 1973).

[4] J.L. Tonry, in {\it Texas/PASCOS 92: Relativistic Astrophysics
and Particle Cosmology}, edited by C.W. Akerlof and M.A. Srednicki
(Ann. NY Acad. Sci. {\bf688}, 113 (1993)).

[5] K.R. Lang, {\it Astrophysical Formulae\/} (Springer-Verlag,
Berlin, 1974), section 1.32.

[6] Particle Data Group, {\it Review of Particle Physics\/} (Eur.
Phys. J. C {\bf3}, 70 (1998)).

[7] The proof given in the appendix in Ref. [8] is conducted in
terms of two complex and conjugate variables, but it proceeds
exactly in the same way as when we substitute our real variables $
u $ and $ v $ instead. The lack of the square power at the last
term in the expression for $ \tau'\left(\xi'\right) $, after Ref.
[8], is a misprint only and therefore does not affect the result.

[8] P. Szekeres, Commun. Math. Phys. {\bf41}, 55 (1975).

[9] In the heliosphere the smallest values of the magnetic field
are observed as fluctuations ($ \pm $ sense of components) of a
few $ \mu $G [10]. The interstellar magnetic field is estimated at
$\sim1 $ $\mu$G [5,11].

[10] NSSDC COHOWeb Data Explorer Results, {\it Plot for Ulysses
data from 901025 to 971231\/}
(http://nssdc.gsfc.nasa.gov/space/).

[11] P.J.E. Peebles, {\it Principles of Physical Cosmology\/}
(Princeton University Press, Princeton NJ, 1993), pp. 149 and
654.

[12] S. Bajtlik, private communication.

[13] P.P. Kronberg, Rep. Prog. Phys. {\bf57}, 325 (1994).

[14] M. Lemoine {\it et al}., Astrophys. J. {\bf486}, L115 (1997).

[15] B. Bertotti, Phys. Rev. {\bf 116}, 1331 (1959).

[16] I. Robinson, Bull. Acad. Polon. Sci., Ser. Math. Astr.
Phys. {\bf 7}, 351 (1959).

\end{document}